\documentclass[conference]{IEEEtran}
\usepackage{tikz, pgfplots}
\usepackage{multirow, booktabs, url, tabularray}
\usepackage{amssymb, amsmath}
\usepackage{colortbl}
\pgfplotsset{compat=1.18}
\usepackage{url}
\usepackage{array}
\newcolumntype{L}[1]{>{\raggedright\let\newline\\\arraybackslash\hspace{0pt}}m{#1}}
\newcolumntype{C}[1]{>{\centering\let\newline\\\arraybackslash\hspace{0pt}}m{#1}}
\newcolumntype{R}[1]{>{\raggedleft\let\newline\\\arraybackslash\hspace{0pt}}m{#1}}
\pgfplotsset{tick label style={/pgf/number format/fixed,/pgf/number format/precision=2},scaled y ticks=false}
\usepackage[noadjust]{cite}

\title{Boosting GUI Prototyping with Diffusion Models}
\author{
Jialiang Wei$^*$, Anne-Lise Courbis$^*$, Thomas Lambolais$^*$, \\Binbin Xu$^*$, Pierre Louis Bernard$^{**}$ and Gérard Dray$^*$
\\[6pt]
$^{*}$: {\normalsize EuroMov Digital Health in Motion, Univ Montpellier, IMT Mines Ales, Ales, France}\\
$^{**}$: {\normalsize EuroMov Digital Health in Motion, Univ Montpellier, IMT Mines Ales, Montpellier, France}\\
\texttt{$^{*}$: {\small firstname.lastname@mines-ales.fr} $^{**}$: {\small firstname.lastname@umontpellier.fr}}
}

\begin{document}
\maketitle

\begin{abstract}
GUI (graphical user interface) prototyping is a widely-used technique in requirements engineering for gathering and refining requirements, reducing development risks and increasing stakeholder engagement. 
However, GUI prototyping can be a time-consuming and costly process.
In recent years, deep learning models such as Stable Diffusion have emerged as a powerful text-to-image tool capable of generating detailed images based on text prompts.
In this paper, we propose UI-Diffuser, an approach that leverages Stable Diffusion to generate mobile UIs through simple textual descriptions and UI components.
Preliminary results show that UI-Diffuser provides an efficient and cost-effective way to generate mobile GUI designs while reducing the need for extensive prototyping efforts. This approach has the potential to significantly improve the speed and efficiency of GUI prototyping in requirements engineering.
\end{abstract}

\section{Introduction}
The exponential growth of mobile technology and the increasing dependence on mobile devices for various daily activities have significantly influenced the design and development of mobile applications. 
The mobile GUI (graphical user interface) plays a critical role in mobile applications since it is the primary means of interaction between users and their devices.
A well-designed mobile UI can significantly enhance the user experience, making it simpler for users to navigate and achieve their desired tasks, resulting in increased user engagement and retention \cite{Chen2021,Hassan2020}.
Additionally, an engaging and user-friendly GUI can set an application apart from its competitors and increase its chances of success in the highly competitive mobile app market \cite{Moran2018}.
As the competition among mobile apps intensifies, it's critical for developers to create innovative and user-friendly GUI to meet users' evolving expectations. 

GUI prototyping is a crucial technique that allows developers to create an initial version of a GUI design, assess its effectiveness, and refine it based on feedback from stakeholders. 
This technique is highly valuable in the context of requirements engineering, as it can help refine requirements, reduce development risks, and promote stakeholders' engagement~\cite{Silva2015}. 
Despite the benefits of GUI prototying, it can be a time-consuming and costly process \cite{Shams-ul-Arif2010}.

To enhance the GUI prototyping process, various approaches have been proposed, including the use of established tools that provide basic components and templates for creating GUIs. 
The industry has widely embraced commercial tools like Figma \cite{Figma}, InVision Studio \cite{InVision}, Adobe XD \cite{Adobe}, Moqups \cite{Moqups}, Sketch \cite{Sketch}, as well as open source tools like Pencil Project \cite{Pencil} to  streamline the prototyping process.
In recent literature, several GUI search and retrieval approaches have been proposed \cite{Behrang2018, Huang2019, Chen2020, Li2021, Bunian2021, Bernal-Cardenas2019, Feng2022, Kolthoff2023}.
These approaches aim to provide users with the ability to search for inspiration from existing designs and reuse them to streamline the GUI prototyping process, which can reduce the time and effort spent on creating new designs.

\begin{figure}[!htb]
\centerline{
    \includegraphics[width=0.45\textwidth]{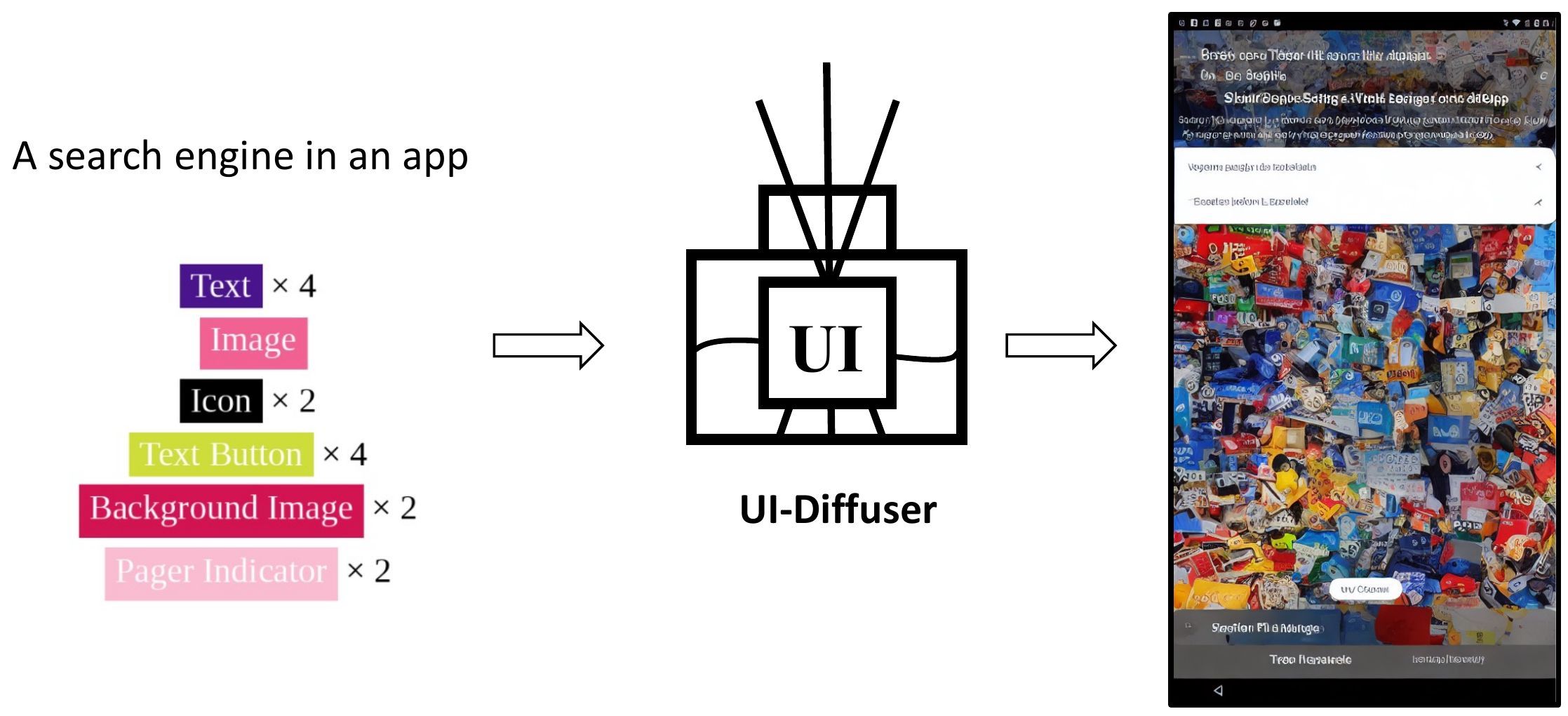}}
    \caption{UI image generation with UI-Diffuser}
    \label{fig:image-generation}
\end{figure}

Recently, deep learning-based text-to-image models have emerged as a promising approach to generate highly detailed and structured images based on text descriptions \cite{Ramesh2022,Rombach2022,Saharia2022,Nichol2021}.
Among these models, Stable Diffusion \cite{Rombach2022} has shown impressive results in generating high-quality images from textual input.
In this study, we propose a novel approach called UI-Diffuser that leverages the Stable Diffusion model to generate mobile UI designs through simple text prompts and UI components, as illustrated in Figure \ref{fig:image-generation}.
The preliminary results of UI-Diffuser show its potential to enhance the effectiveness of mobile UI design and to decrease time consumption for prototyping, leading to cost reductions for this phase.

The rest of the paper is organized as follows: 
Section \ref{sec:background} presents the background of this research.
Section \ref{sec:approach} details the framework of UI-Diffuser.
Section \ref{sec:demo} shows the samples of generated UI images and discusses their limitations.
Section \ref{sec:conclusion} concludes the paper and outlines the future directions.

\section{Background}\label{sec:background}
\subsection{GUI Prototyping}
To streamline and enhance the process of GUI prototyping, numerous strategies have been previously introduced.

Prototyping tools that provides basic components and templates are widely used in practice, such as Figma~\cite{Figma}, InVision Studio~\cite{InVision}, Adobe XD~\cite{Adobe}, Moqups~\cite{Moqups}, Sketch~\cite{Sketch}, and Pencil Project~\cite{Pencil}.
However, utilizing these tools effectively requires users to possess design experience.

Some researchers introduced GUI retrieval approaches that take the sketch (GUIfetch~\cite{Behrang2018}, SWiRE~\cite{Huang2019}, Wireframe-Based UI Design Search of Chen et al.~\cite{Chen2020}) or the screenshot (Screen2vec~\cite{Li2021}, VINS~\cite{Bunian2021}) as input and find designs that are similar to the input.
Although useful, these methods require an initial rudimentary GUI prototype.
Moreover, while these strategies primarily aim to support the GUI design's prototyping by presenting alternative designs, they are not ideal for interactive GUI prototyping during requirements elicitation.

GUI search engines, such as Guigle~\cite{Bernal-Cardenas2019}, Gallery D.C.~\cite{Feng2022}, and RaWi~\cite{Kolthoff2023}, allow users to search for existing GUI designs and components using textual queries. 
Wei et al. \cite{Wei2022REW} have put forth an proposition of combining app features with app UIs.
While these approaches can serve as a source of inspiration and guidance, blindly copying existing designs without proper attribution or permission can result in legal issues such as copyright infringement.

GUIGAN, as introduced by Zhao et al. \cite{Zhao2021}, leverages previously collected GUI components from existing mobile app to compose new designs. 
These new composite designs not only comply with accepted standards of GUI structure, but also cater to consumer aesthetics.
However, it is important to note that this approach offers limited control over the generation process.

In summary, these existing approaches fall short in providing support to analysts during the requirements elicitation phase through GUI prototyping.

\subsection{Image synthesis}
Image synthesis is a process of creating new images from diverse forms of image descriptions, including textual descriptions, sketch images, noise, and others.

In 2014, Goodfellow et al. \cite{Goodfellow2020} introduced Generative Adversarial Network (GAN) as a means of generating realistic image.
GANs consist of two deep neural networks: a generator and a discriminator. 
The generator creates new images to deceive the discriminator, which aims to distinguish real from fake images. 
Both networks are trained simultaneously and this process continues until the generator produces images that are indistinguishable from real images.
However, the lack of diversity and the challenges associated with training GANs limit their scalability and hinder their applicability to novel domains \cite{Dhariwal2021}.

Diffusion models (DMs) \cite{Ho2020} are neural networks trained to denoise images blurred with Gaussian noise by learning to reverse the diffusion process.
Recent studies \cite{Ho2020} have demonstrated that DMs are capable of generating high-quality images, and possess desirable attributes such as distribution coverage, a stable training objective, and scalability.
Several companies recently released their image synthesis tools based on DMs like DALL$\cdot$E 2~\cite{Ramesh2022}, Midjourney~\cite{Midjourney}, Stable Diffusion~\cite{Rombach2022}, and DreamBooth~\cite{Ruiz2022}.
Despite their usefulness in generating images, existing tools for image synthesis generally lack efficacy in generating UIs.
Since 2023, some researchers utilize DMs for the generation of UI layout~\cite{Hui2023,Cheng2023,Levi2023,Inoue2023}.
However, to the best of our knowledge, no DM-based models have been developed specifically for generating UIs.

\section{UI-Diffuser}\label{sec:approach}
\begin{figure}[!htb]
\centerline{
    \includegraphics[width=0.45\textwidth]{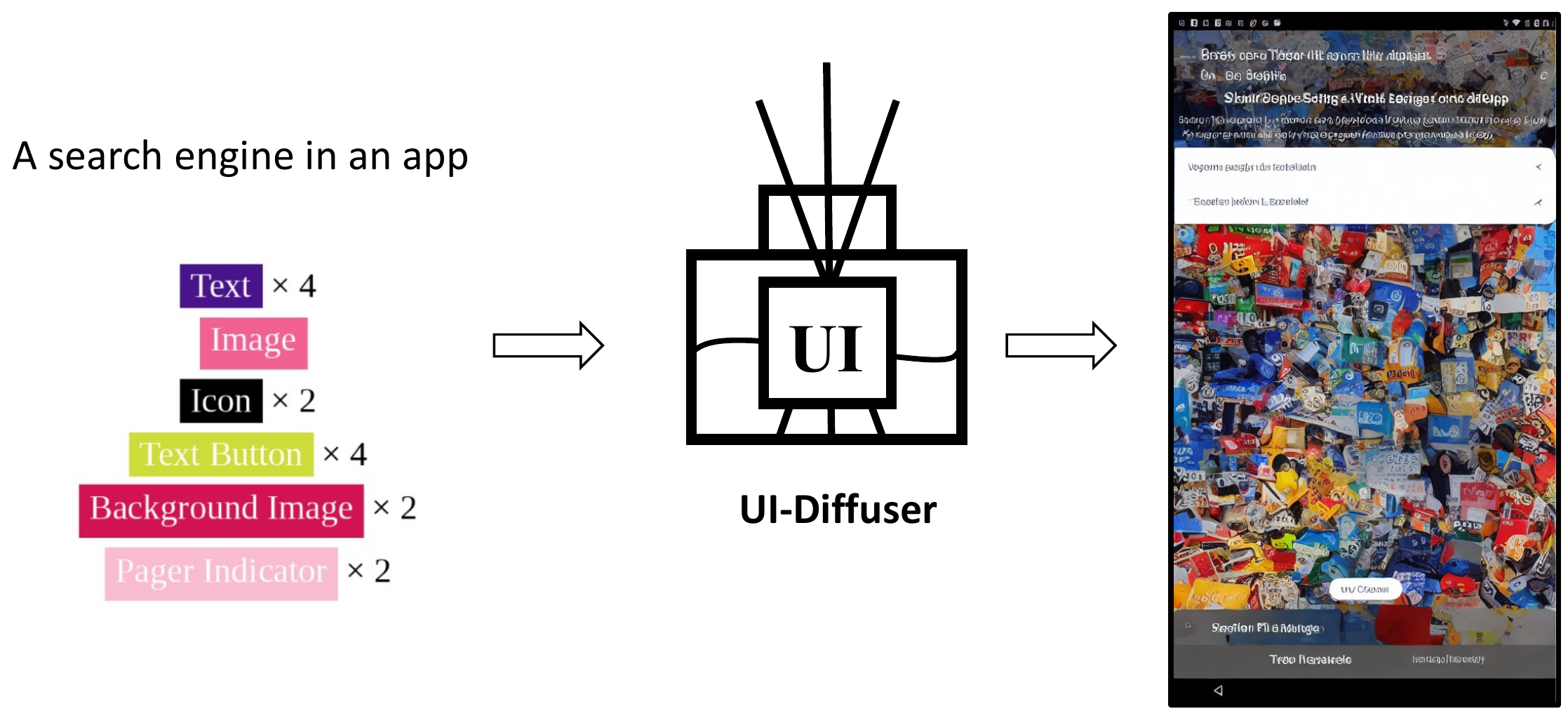}}
    \caption{Overview of UI-Diffuser}
    \label{fig:overview}
\end{figure}

UI-Diffuser is a novel approach that facilites requirements engineers in rapidly prototyping mobile app UIs through a two-step process (cf. Figure~\ref{fig:overview}). 
In the first step, UI-Diffuser takes input UI components, such as text, buttons, and images, and generates a layout that considers the arrangement of these components (see Section \ref{sec:layout-generation}). 
In the second step, the generated layout is used to complete a mobile UI image based on the textual description provided by the user (see Section \ref{sec:ui-generation}). 
In the subsequent subsections, we will describe each step of UI-Diffuser in detail.

\subsection{Layout Generation}\label{sec:layout-generation}
The objective of layout generation is to create realistic graphic scenes that consist of diverse components with varying attributes, such as category, size, position, and between-component relationships \cite{Hui2023}.
This task is critical for simplifying graphic design tasks, especially for structured scenes like documents and user interfaces.

To achieve this goal, we employ LayoutDM \cite{Inoue2023} in this study.
LayoutDM builds on the discrete-state space diffusion models \cite{Gu2022, JacobAustin2021} and has been trained on the Rico dataset \cite{Deka2017} -- a dataset of user interface designs for mobile applications containing 25 categories of UI components, such as text button, toolbar, and icon.
We will detail the Rico dataset in Section \ref{sec:dataset}.
LayoutDM allows the generation of UI layout with given conditions, such as a list of components.
Figure \ref{fig:overview} illustrates the generation of layout with a given component list. 

\subsection{UI Generation}\label{sec:ui-generation}
Given the generated layout and a textual description, UI-Diffuser is able to generate UI images fitting the layout and the description.

\subsubsection{Architecture}
\begin{figure}[!htb]
\centerline{
    \includegraphics[width=0.48\textwidth]{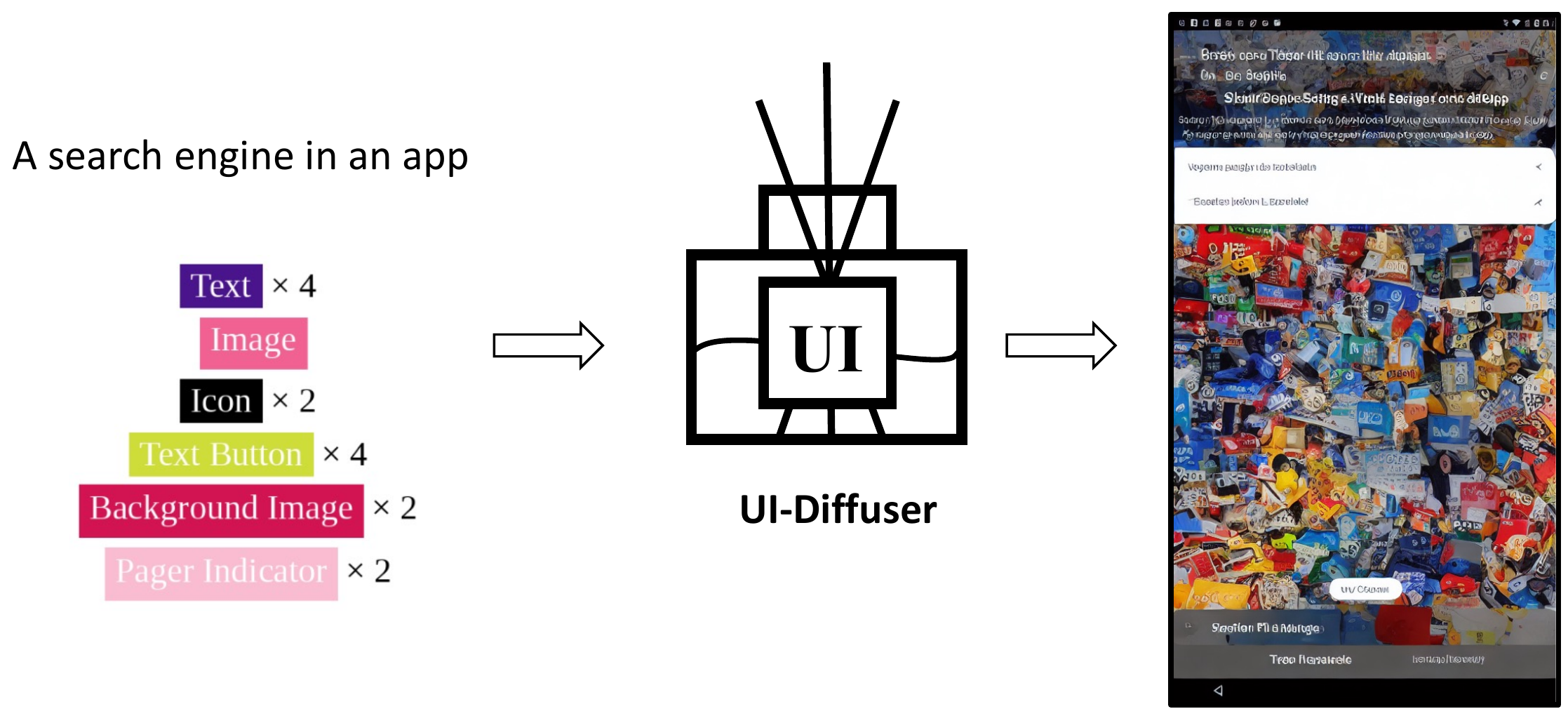}}
    \caption{Overview of UI generation model}
    \label{fig:sd}
\end{figure}

To generate UI images from layout and description, we utilized Stable Diffusion \cite{Rombach2022} augmented with ControlNet \cite{Zhang2023}.
The proposed model workflow is illustrated in Figure \ref{fig:sd}.
It contains four components: Text Encoder, Image Information Creator, Image Decoder, and ControlNet.

The Text Encoder transforms raw text inputs into numerical vectors called token embeddings.
CLIP \cite{Radford2021} is a pre-trained model that has been trained on a large-scale corpus of text and images using a contrastive learning approach.
CLIP is employed as text encoder due to its remarkable performance in encoding both text and images into a shared latent space.

The Image Information Creator is responsible for generating image embeddings based on the given token embeddings, which are then used by the Image Decoder to produce the final image.
The \textit{diffusion process} occurs inside this component, in a step-by-step fashion.  
Starting from a noisy image, each step of the \textit{diffusion process} adds more relevant information that aligns with the input text.
UNet \cite{Ronneberger2015} is used in this component.
During the pre-training, a large number of images are blurred with Gaussian noise and the UNet is trained by denoising the image.

To support additional input conditions, such as the layout image in our case, ControlNet~\cite{Zhang2023} is integrated with UNet.
ControlNet is an end-to-end neural network architecture that enables the control of large image diffusion models, like Stable Diffusion, to learn task-specific input conditions.
ControlNet clones the weights of the Image Information Creator into a ``trainable copy''.
The original Image Information Creator preserves the network capability learned from billions of images during the pre-training, while the ``trainable copy'' is trained on the Rico datasets to learn the conditional control.
The ``trainable copy'' and the Image Information Creator are connected with a unique type of convolution layer called ``zero convolution''.
We will train the ControlNet during the fine-tuning.

Finally, the Image Decoder generates images using the image embeddings.
Variational Autoencoder (VAE)~\cite{Kingma2014} is a type of generative model consisting of an encoder network, which maps input images into a lower-dimensional latent space, and a decoder network, which maps the latent space back to the original images space.
The VAE encoder-decoder pair have been pre-trained on a large number of images to accurately reconstruct input images.
The decoder network of VAE is used as the Image Decoder.

\subsubsection{Dataset for fine-tuning}\label{sec:dataset}
The fine-tuning of ControlNet \cite{Zhang2023} requires a dataset including input images, conditioning images and text prompts.
In the context of UI generation, this dataset should consist of UI screenshots, wireframes that depict the page layout, and textual descriptions of the screenshots. 
To this end, we leveraged the Rico dataset \cite{Deka2017}, which provides the first two elements: the screenshots and wireframes. 
And we generated the textual descriptions of these screenshots using XUI \cite{Leiva2023}.

The Rico dataset \cite{Deka2017} is one of the largest mobile app design datasets to date, encompassing design data from more than 9.3k Android applications across 27 categories. 
This dataset provides access to the visual, textual, structural, and interactive design properties of more than 66k distinct UI screens. 
In this work, we utilized the screenshots, wireframes, and hierarchies of the Rico dataset.

\begin{figure}[!htb]
\centerline{
    \includegraphics[width=0.3\textwidth]{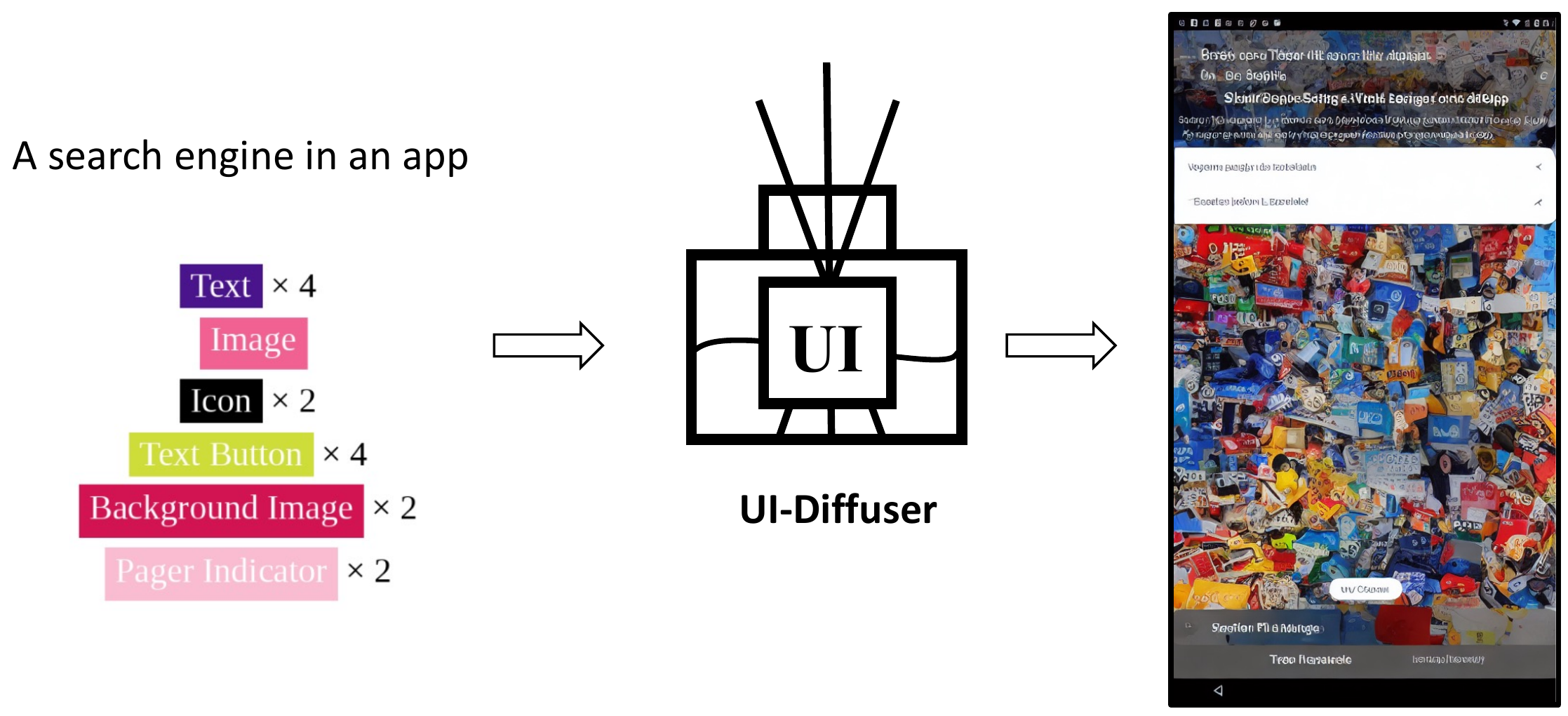}}
    \caption{Example of a screenshot and its wireframe from Rico dataset}
    \label{fig:screenshot}
\end{figure}

To prepare the Rico dataset for fine-tuning the Stable Diffusion model, a preprocessing pipeline was applied to its screenshots and wireframes. The pipeline includes the following steps:
\begin{itemize}
\item remove the screenshots and wireframes on landscape.
For this preliminary work, we focus on portrait UI design.
\item resize the screenshots and wireframes to 288x512.
The original size of the images is 1080x1920 or 540x960.
However, the Stable Diffusion model we use accept only images with height/width less than 512 as its input during the training process.
\end{itemize}

\begin{table*}[]
\caption{Samples of generated UI images}
\centering
\begin{tblr}{h{2cm}|Q[m,c]|Q[m,c]|Q[m,c]|Q[m,c]|Q[m,c]|Q[m,c]}
\hline[1pt]
\SetCell[r=1,c=1]{m,c}\textbf{Components and Descriptions} & \SetCell[r=1,c=1]{m,c}\textbf{Generated Layouts} & \SetCell[r=1,c=5]{m,c} {\textbf{Generated UIs}}\\ 
\hline
\includegraphics[width=18mm]{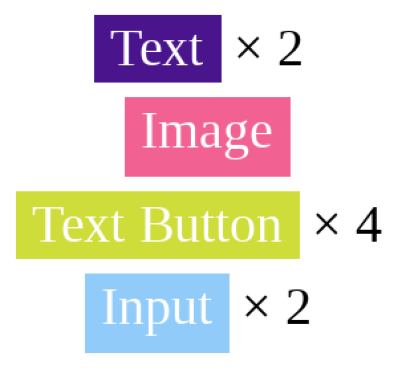} 
A login page with input fields.
& 
\fbox{\includegraphics[scale=0.18]{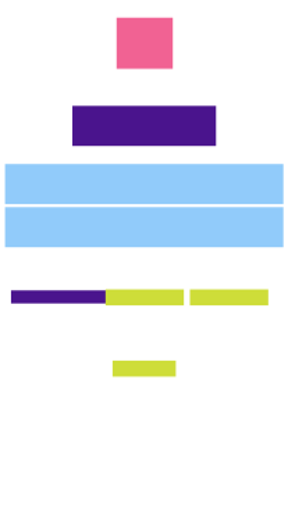}} & 
\fbox{\includegraphics[scale=0.18]{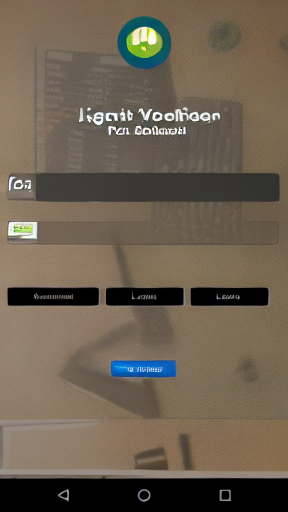}} & 
\fbox{\includegraphics[scale=0.18]{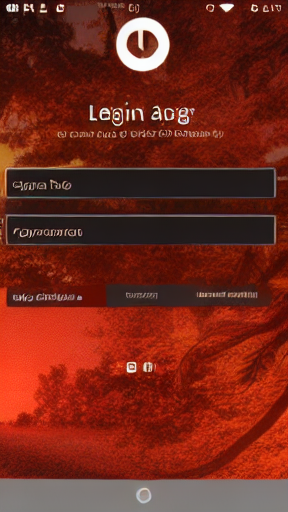}} & 
\fbox{\includegraphics[scale=0.18]{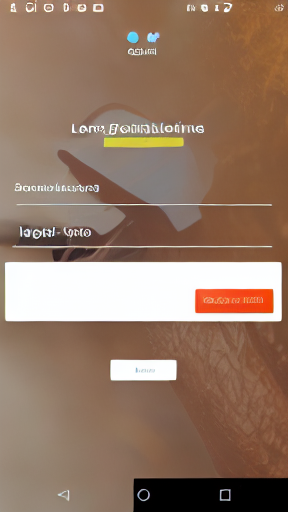}} &
\fbox{\includegraphics[scale=0.18]{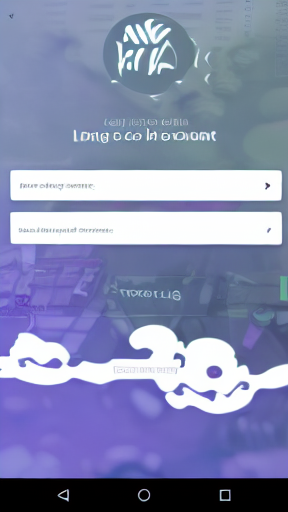}} &
\fbox{\includegraphics[scale=0.18]{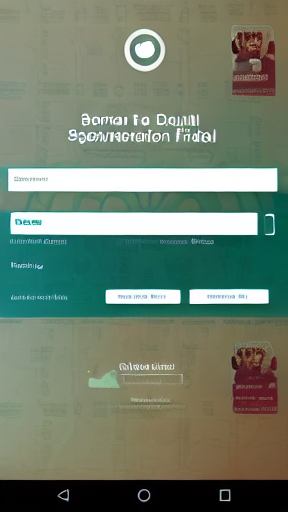}}\\
\hline
\includegraphics[width=15mm]{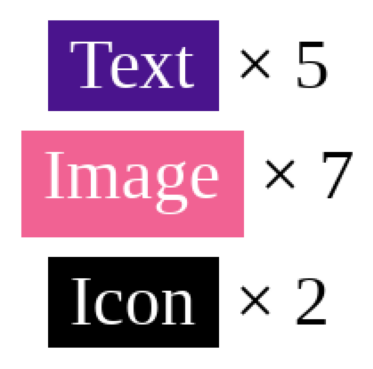} 
A tutorial app having text components.
& 
\fbox{\includegraphics[scale=0.18]{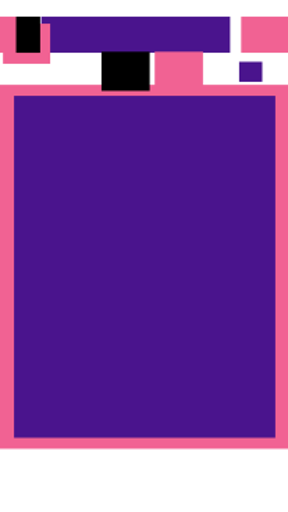}} & 
\fbox{\includegraphics[scale=0.18]{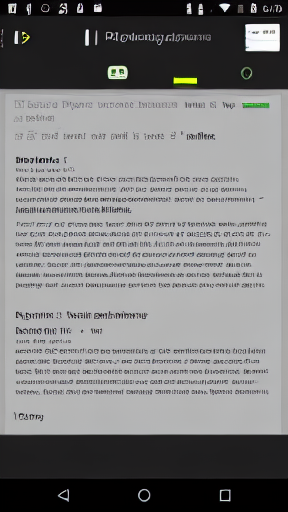}} & 
\fbox{\includegraphics[scale=0.18]{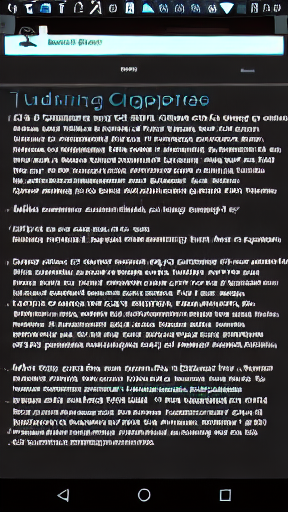}} & 
\fbox{\includegraphics[scale=0.18]{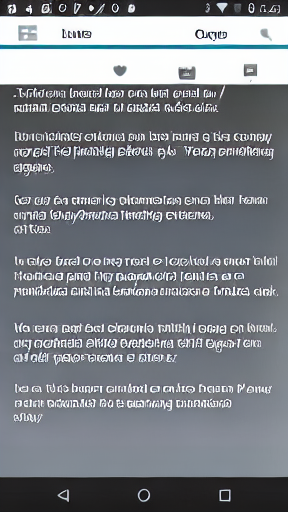}} &
\fbox{\includegraphics[scale=0.18]{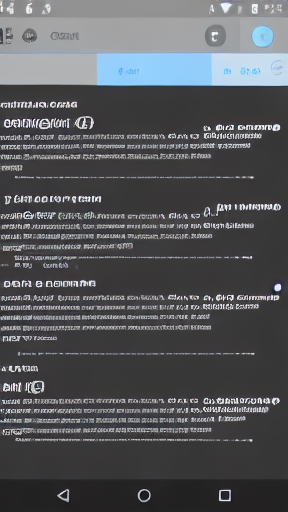}} &
\fbox{\includegraphics[scale=0.18]{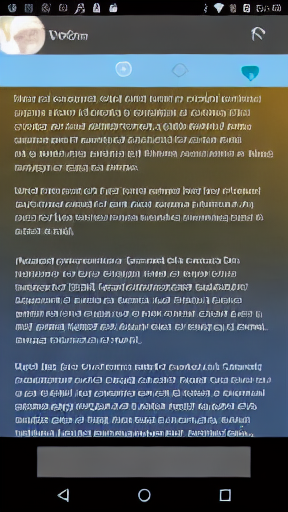}}\\
\hline
\includegraphics[width=20mm]{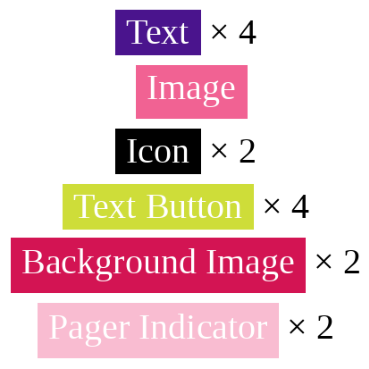}
A gallery page of an app.
&
\fbox{\includegraphics[scale=0.18]{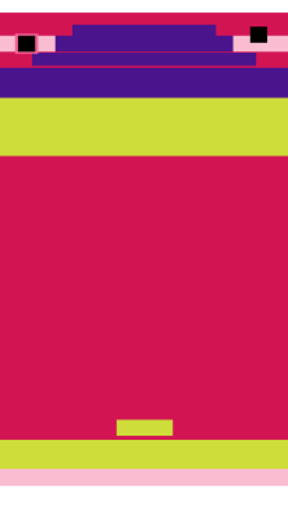}} & 
\fbox{\includegraphics[scale=0.18]{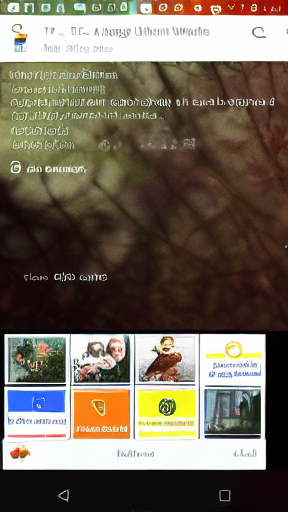}} & 
\fbox{\includegraphics[scale=0.18]{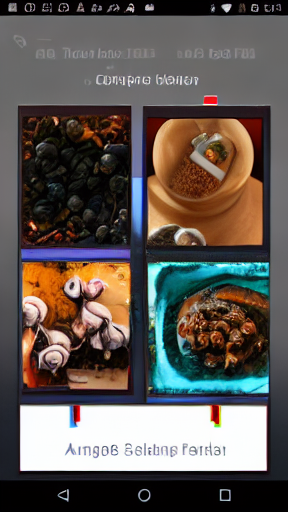}} &
\fbox{\includegraphics[scale=0.18]{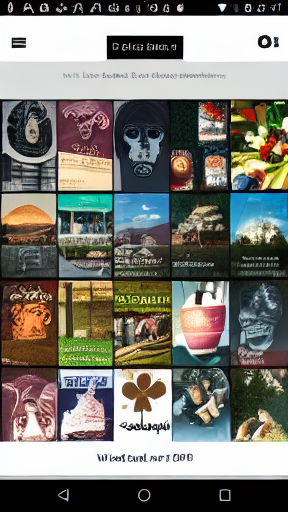}} &
\fbox{\includegraphics[scale=0.18]{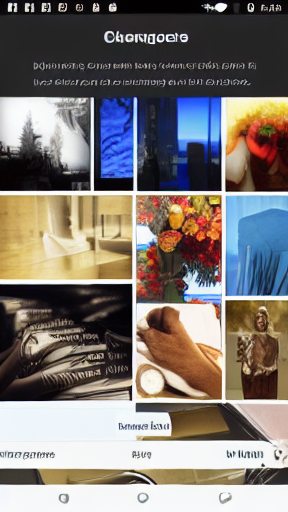}} & 
\fbox{\includegraphics[scale=0.18]{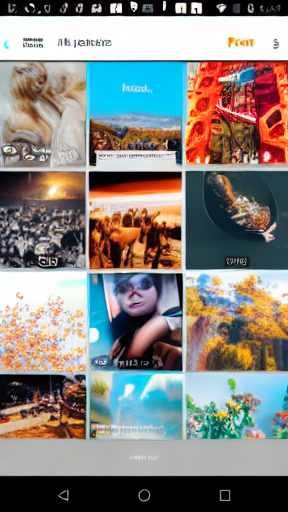}}\\
\hline
\includegraphics[width=18mm]{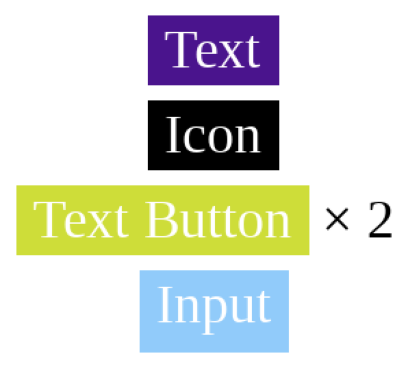} 
A maps app.
& 
\fbox{\includegraphics[scale=0.18]{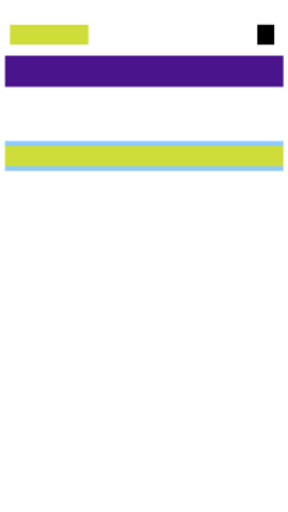}} & 
\fbox{\includegraphics[scale=0.18]{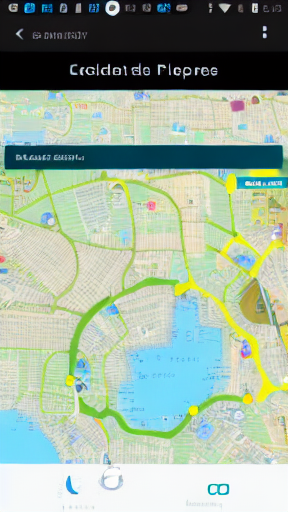}} & 
\fbox{\includegraphics[scale=0.18]{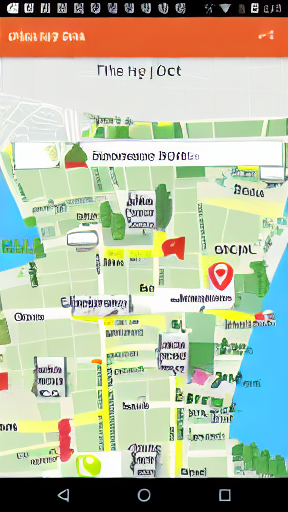}} &
\fbox{\includegraphics[scale=0.18]{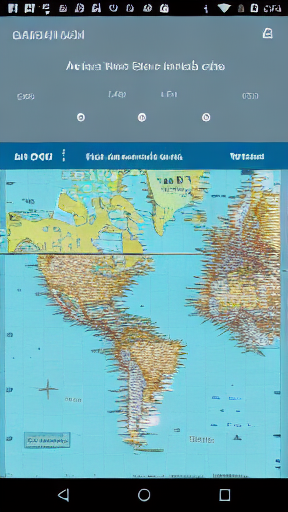}} &
\fbox{\includegraphics[scale=0.18]{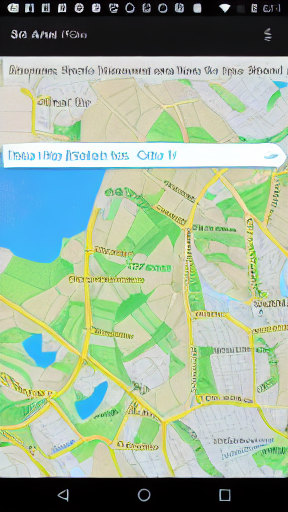}} &
\fbox{\includegraphics[scale=0.18]{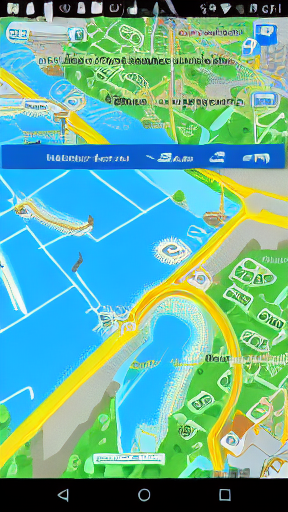}}\\
\hline
\includegraphics[width=18mm]{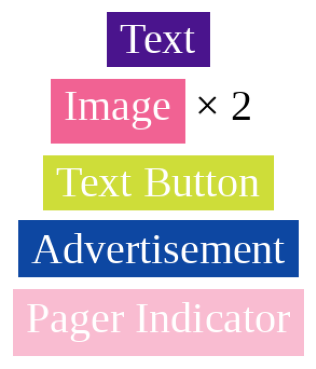} 
A mediaplayer app.
& 
\fbox{\includegraphics[scale=0.18]{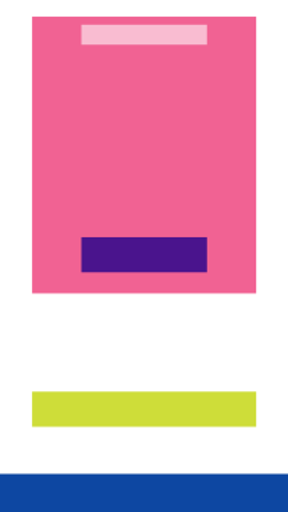}} & 
\fbox{\includegraphics[scale=0.18]{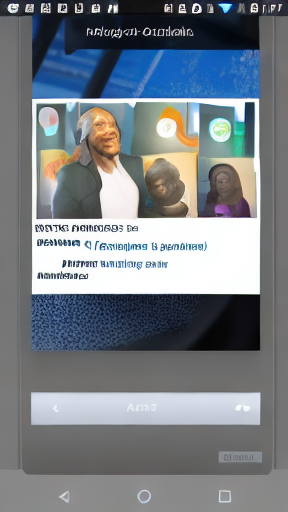}} & 
\fbox{\includegraphics[scale=0.18]{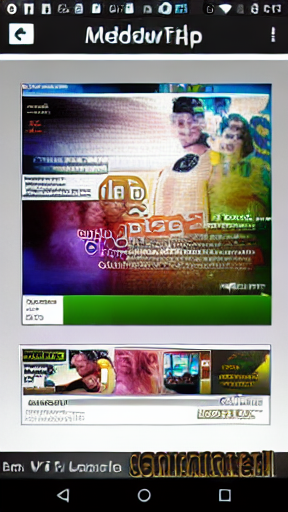}} &
\fbox{\includegraphics[scale=0.18]{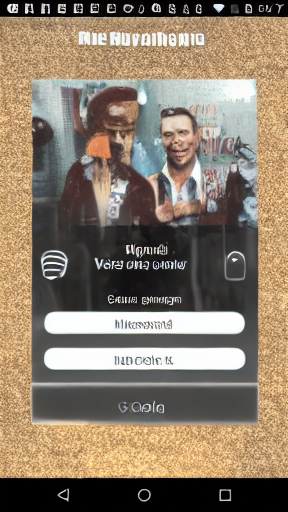}} &
\fbox{\includegraphics[scale=0.18]{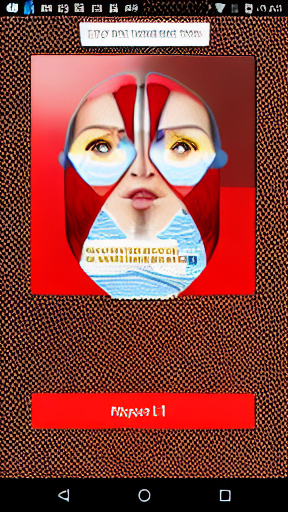}} &
\fbox{\includegraphics[scale=0.18]{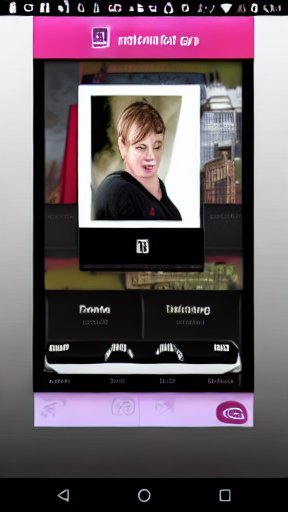}} & \\
\hline
\includegraphics[width=20mm]{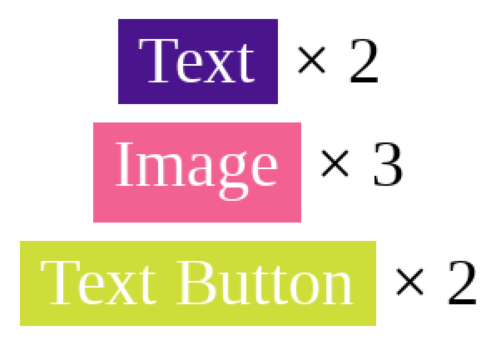} 
A profile app with a big image.
& 
\fbox{\includegraphics[scale=0.18]{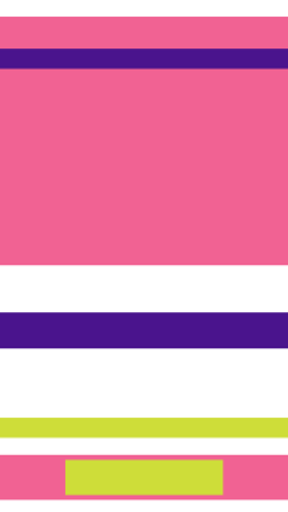}} & 
\fbox{\includegraphics[scale=0.18]{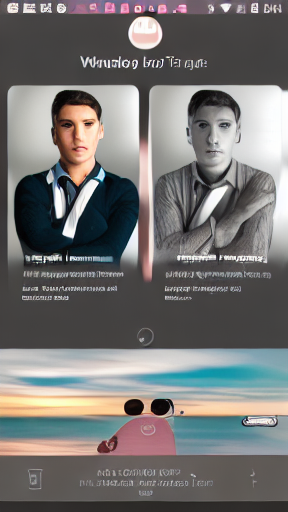}} & 
\fbox{\includegraphics[scale=0.18]{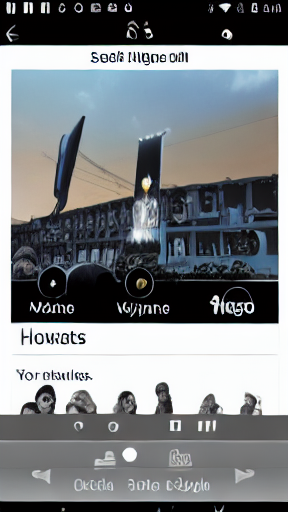}} &
\fbox{\includegraphics[scale=0.18]{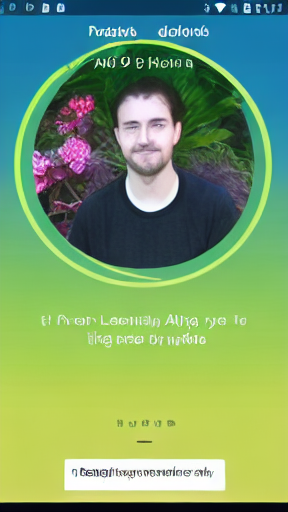}} &
\fbox{\includegraphics[scale=0.18]{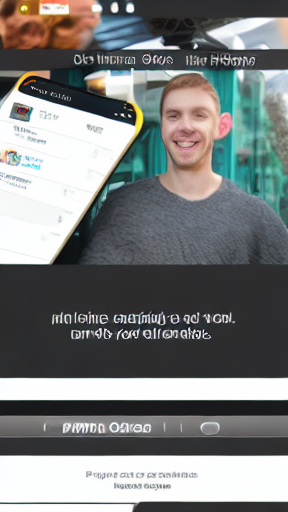}} &
\fbox{\includegraphics[scale=0.18]{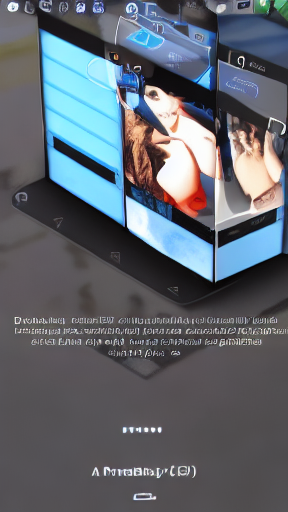}} &\\
\hline[1pt]
\end{tblr}
\label{tab:samples}
\end{table*}

The Rico dataset does not provide the textual description of the screenshots.
For this reason, we used XUI \cite{Leiva2023} model to generate the descriptions.
XUI is a tool to generate automatically informative description of a given UI screenshot.
It takes the screenshots, wireframes, and hierarchies of the UIs as input, and generate their natural language descriptions.
For instance, the generated description of Figure \ref{fig:screenshot} is ``\textit{That screen maybe is a list screen. You may see a list of elements, typically arranged in rows. You may notice a text ubicated at the center area.}''.
During the training, we randomly replace 50\% text prompts with a default prompt (``\textit{A nice screenshot of a mobile app}''). 
This facilitates ControlNet's capability to better recognize semantic contents from the wireframes.
This is primarily because, in cases where the prompt lacks sufficient information for the Stable Diffusion model, the model may rely more heavily on the semantics of the input wireframes to compensate for the lack of meaningful guidance from the prompt.

The processed screenshots and wireframes from Rico dataset and the generated textual descriptions of XUI are then used for the subsequent fine-tuning.

\subsubsection{Fine-tuning procedures}
During the fine-tuning procedures, the parameters of Text Encoder, the Image Information Creator, and the Image Decoder are all locked.
The update is only applied to the parameters of ControlNet.

Our model are implemented with the Diffusers library\footnote{https://github.com/huggingface/diffusers}.
We employed the ``{runwayml/stable-diffusion-v1-5}'' checkpoint\footnote{https://huggingface.co/runwayml/stable-diffusion-v1-5} and the ``lllyasviel/sd-controlnet-seg'' checkpoint\footnote{https://huggingface.co/lllyasviel/sd-controlnet-seg} from HuggingFace.
We trained the model for 1 epoch on the processed Rico dataset with a batch size of 4, equivalent to about 16,000 steps.
We use AdamW for optimization with a learning rate of $1e^{-5}$.
The training was performed using Nvidia Tesla T4 with 16GB VRAM.

\section{Demo and Discussion} \label{sec:demo}
Table \ref{tab:samples} illustrates some samples of UI images generated by UI-diffuser.
Given UI components and a brief description, UI-Diffuser generates UIs with various designs that roughly meet the requirements.
UI-Diffuser can produce a UI image within a matter of seconds. 
Compared to traditional prototyping, UI-Diffuser can generate GUI prototypes at a much quicker rate.

At first glance, the UI images generated by UI-diffuser appear to be of high quality. 
However, upon closer examination, some details are missing.
It's important to note that the generated UIs may not always conform to the components category, as illustrated by the five UIs in the fifth row of Table \ref{tab:samples}, where the "advertisement" component at the bottom of the layout is disregarded. 
Additionally, certain generated UIs may not meet aesthetic standards.

Consequently, the current UIs produced by UI-Diffuser may be more suitable for inspiring UI designers than serving as fully functional UI prototypes.

\section{Conclusion and Roadmap}\label{sec:conclusion}
This paper presents UI-Diffuser, a GUI prototyping tool that utilizes layout components and simple text prompts to produce mobile UI designs.
Through a demo, we demonstrate that UI-Diffuser can generate UI images that align the given components and textual descriptions, highlighting the potential advantages of UI-Diffuser in GUI prototyping.

To advance this research, we intend to perform a comprehensive evaluation of the UI-Diffuser by investigating three critical factors: 
the aesthetics of the generated UIs, their compatibility with UI components, and their compatibility with textual descriptions.
The aesthetics evaluation will be carried out manually. 
We shall rate the aesthetics based on a predetermined set of criteria.
As for the compatibility with UI components, we will manually assess the number of correctly generated UI components.
Finally, we will use CLIPScore \cite{Hessel2021} to calculate the compatibility of the descriptions and their generated UIs.

Moreover, we propose enhancing UI-Diffuser from three aspects:

\textit{Developping a dataset with high-quality screenshot descriptions:}
the quality of the image descriptions within a training dataset has a significant impact on the performance of Stable Diffusion.
In this work, we used XUI \cite{Leiva2023} to generate screenshots' descriptions.
Although XUI is a valuable tool for generating descriptions of screenshots, it only categorizes screenshots into roughly 20 categories, which is insufficient in depicting the numerous functions of modern mobile apps.
Moreover, the descriptions generated by XUI lack sufficient detail in describing the UI components.
To address these limitations, we plan to investigate alternative UI image captioning tools that can generate more comprehensive UI descriptions.

\textit{Cropping components from generated UIs:} 
while generated UI images can inspire requirements engineers in GUI prototyping, they are usually not editable or directly reusable. 
To overcome this limitation, we take inspiration from Kolthoff et al. \cite{Kolthoff2023} and propose cropping each component of the generated UI image based on its absolute position in the generated layout image. 
The GUI components may overlap with each other, leading to a blank space of the lower component when cropping the top component. 
In such cases, the blank space can be filled with the top-ranked RGB color from the color histogram of the lower component.
The cropped components can be then reused in further prototyping.

\textit{Generating code from generated UIs:}
the ability to generate code from UI designs can significantly accelerate the development of application prototypes.
As the layout images generated by UI-Diffuser contain the components' category as well as their size and position, it is possible to generate corresponding code \cite{DeSouzaBaule2020}.
We intend to develop a GUI code generator that compromises two steps: 
(i) extract location and size of each components from generated layout image, and the style from generated UI image,
(ii) generate GUI code for each kind of components according to its attributes.

We believe that exploring the potential of Diffusion Models for generating UIs is a promising research direction, as it can significantly improve the speed and efficiency of GUI prototyping in requirements engineering.

\bibliographystyle{IEEEtran}
\bibliography{ref}
\end{document}